\documentclass[12pt]{article}
\usepackage{epsfig}
\usepackage{amsmath}
\usepackage{natbib}

\usepackage{amsfonts}
\usepackage{rotating}
\usepackage{simplemargins}
\setallmargins{1in}

\begin{document}

\begin{center}
\begin{Large}
A Changepoint Detection Method for Profile Variance
\end{Large}

\vspace{.2in}

Vladimir J. Geneus\footnote{Corresponding author: vgeneus@stat.fsu.edu}, Eric Chicken \\
Department of Statistics \\
Florida State University \\

\vspace{.2in}

Jordan Cuevas \\
Johnson and Johnson Vision Care \\
Jacksonville, Florida\\

\vspace{.2in}

Joseph J. Pignatiello, Jr.\\
Department of Operational Sciences \\
Air Force Institute of Technology \\

\end{center}

\vspace{.2in}

\flushbottom 

\begin{small}
\abstract{A wavelet-based changepoint method is proposed that determines when the variability of the noise in a sequence of functional profiles goes out-of-control from a known, fixed value. The functional portion of the profiles are allowed to come from a large class of functions and may vary from profile to profile.  The proposed method makes use of the orthogonal properties of wavelet projections to accurately and efficiently monitor the level of noise from one profile to the next.  Several alternative implementations of the estimator are compared on a variety of conditions, including allowing the wavelet noise subspace to be substantially contaminated by the profile's functional structure.  The proposed method is shown to be very efficient at detecting when the variability has changed through an extensive simulation study.
\\
\begin{footnotesize}
\end{footnotesize}
}
\end{small}

\thispagestyle{empty}

\section{Introduction}

Traditional statistical process control (SPC) methods have been used to monitor univariate processes for changes in means, standard deviations and many other parameters of interest.  Multivariate control charts are also available to monitor for changes in mean vectors and covariance matrices.  However, data collection technology has improved to the point that massive amounts of data are often readily available and exist in complex data structures that may not be readily or best addressed by these traditional uni- and multivariate methods.  For example, a ``single observation" of an in-control process might consist of a functional response in the form of $n$ pairs of $(x, y)$ data that can be described by $y = f (x) +\epsilon$ where $f$ is a known function and $\epsilon$ is random noise with mean zero and standard deviation $\sigma$. Thus, an observation on the process is a realization of a dependent variable $y$ at $n$ values of $x$.

Such data structures or relationships between $y$ and $x$ are
called profiles. Examples of profiles include calibration
curves in chemical processing \cite{stover}, oxide thickness
across wafer surfaces in semiconductor manufacturing
\cite{gardner}, the stamping force as a function of crank arm
angle in a steel stamping operation \cite{jinshi2}, radius measurements as a function of a turning process \cite{colosimo} and radar reflections as a function of angle \cite{chicken}.
In instances such as these, it is desirable to determine as quickly as possible when a change in the
profile has occurred, as this indicates that the process is in an
out-of-control state.

Various control chart methods have been proposed in the literature
for monitoring profiles.  \cite{woodall} provide an excellent overview
of the SPC literature involving profiles, both linear and nonlinear.  Most of these methods are interested in determining when the structure of the function $f$ has changed, rather than the noise.  In fact, many begin by smoothing the data to remove the noise.  For example, \cite{chang} and \cite{shiau} use splines to remove the noise from the observed profiles before characterizing which profiles are different from the nominal in-control profile.  \cite{jinshi1, jinshi2} use wavelets to remove the noise, as well as a tool for both Phase I and Phase II analysis.

In this paper, we focus not on changes in the function $f$ from one profile to the next, but on changes in the variance. \cite{ztw08} proposed a method, known as NEWMA, that would simultaneously monitor for changes in the function along with changes in the noise, under the assumption that the form of the in control function is known. Other profile methods have been proposed that include estimation of the possibly-changing level of noise, but these methods are not monitoring the noise.  \cite{zhang} devised a $\chi^2$ chart for both Phase I and Phase II analysis.  If the profile changes from nominal their chart will note this and classify the corresponding profile as non-nominal, provided that changes in the noise and profile do not cancel each other out.  \cite{ding} implement independent component analysis (ICA) as a tool for Phase I analysis.  They were interested in detecting horizontal mean shifts in the profiles, but make use of an estimate of the noise in their method.

We propose a wavelet-based changepoint method to be used to monitor and detect changes in noise in a Phase II setting.  Changepoint methods have been considered previously in profile methods, though not in conjunction with the use of wavelets except in \cite{chicken} where only functional changes were monitored.  \cite{mahmoud} and \cite{zou} considered changepoint methods in conjunction with linear profiles, while Ding et al.
examined nonlinear profiles.  Ding et al. applied changepoint methods to the components extracted by ICA to detect changes in profiles.  As mentioned above, this was done to detect changes in mean shifts in the functional portion of the profile.

Examples of the use of wavelets for profiles may be found in \cite{fan}, \cite{jinshi1, jinshi2} and \cite{jeong}. Each of these papers used wavelets to detect and classify changes in the structure of $f$ by observing the magnitudes of the coefficients from a profile's wavelet representation.   In contrast, we will use wavelets to monitor for changes in the variance of the profiles.

Unlike Zhang and Albin, we estimate the variance by using within profile, rather than between profile, information.  This allows for efficient estimation of the variance from just a single observed profile.  By observing a sequence of noise estimates from observed profiles, we develop a procedure for determining when the noise has changed and an estimate of the amount of change.

Usually wavelets are used to smooth the function and remove the noise.  This is essentially a data reduction step.  In this paper we take advantage of the ability of wavelets to project a function into a set of subspaces.  In particular, we are interested in a specific projection that consists primarily of noise.  Due to the orthogonality of the wavelet transformation, the normality and independence of the noise in the original profile is carried over into this noise subspace.  Restricting our attention to this subspace allows for efficient estimation of the variance for each profile.

In this paper, we assume that the model for an observed profile $t$ in a sequence is
\begin{equation}
y_i^t=f^t(x_i)+\varepsilon_i^t, \hspace{.1in} i=1, 2, \ldots, n
\label{eq:model}
\end{equation}
where the $y_i^t$ are the observed values of profile $t$ and the noise
is independent normal random variables with variance $\sigma_t^2$.  By monitoring only the wavelet noise subspace, no strong assumptions on the form of $f^t$ are necessary. In fact, the functions $f^t$ can be treated as a nuisance vector and ignored; or monitored alongside the noise chart.

If $f^t$ is considered a nuisance vector, then it should be allowed to vary from one profile to the next without affecting the proposed noise monitoring method.  If $f^t$ is being simultaneously monitored for structure changes, it is important that the noise monitoring methods not confound changes in structure with changes in noise.  It is desirable that a noise monitoring method not depend on $f^t$ so that it can operate independently alongside a separate profile structure monitoring method.  In either case, there is no restriction that the $f^t$ be the same over differing $t$.

If the functions $f^t$ in Equation~\ref{eq:model} were known, then the
variance is easily estimated by examining the difference between the
known profile and the observed data:
\begin{equation}
\varepsilon_i^t=y_i^t-f^t(x_i).
\label{eq:res}
\end{equation}
However, if the functions are not given, then the noise may only be
monitored after they are extracted from Equation~\ref{eq:model}.
If the profile functions were known to follow certain parametric
forms, for example linear or parabolic, estimating $f^t$ could be
done with straightforward statistical techniques such as linear and
non-linear least squares methods.  However, in many
applications of profile monitoring, the forms of the profiles do not
follow easily specified parametric forms.   By using the wavelet noise subspace, we do not require an initial estimate of $f^t$ in order to estimate the noise, as is the case when estimating noise with residual error as in Equation~\ref{eq:res}.

A drawback to estimating the variance $\sigma^2_t$ from the
wavelet-produced noise subspace is that part of the structure of $f^t$ may exist in this space
along with the variance components.
Trying to extract $f^t$ from the noisy model in Equation~\ref{eq:model} via wavelet analysis when this occurs
would result in an poor function estimate: the variance becomes misspecified and leads to oversmoothing the data.  When trying to estimate
the variances, as in this paper, the presence of function structure
alongside the variance components in particular projection spaces
reduces the accuracy of noise estimation.

To overcome this problem, we make use of robust techniques that minimize the effect of this contamination of the noise subspace under certain assumptions on the behavior of $f^t$.  These assumptions do not require that the function $f^t$ be specified,
but instead they impose mild restrictions on the general class of functions
that we can consider.  The estimator we propose effectively monitors profile noise without being unduly affected by profile structure changes.

The remainder of this paper is organized as follows.  Section 2 gives a short background on wavelets.  Section 3 describes the methods we investigate and addresses noise estimation within a single profile.  The proposed changepoint method is detailed in Section 4 and results and conclusions are provided in the final section.

\section{Wavelets and Multiresolution Analysis}

Wavelets are an orthogonal series representation of functions in the
space of square-integrable functions $L_2(R)$. \cite{vidakovic} and
\cite{ogden} offer good introductions to wavelet methods and their
properties. It is common to let $\phi$ and $\psi$ represent the
father and mother wavelet functions, respectively. There are many
choices for these two functions, see \cite{daubechies}. Here, $\phi$
and $\psi$ are chosen to be compactly supported and to generate an
orthonormal basis. Let

$$
\phi_{jk}(x) = 2^{j/2}\phi(2^{j}x-k)
$$
and
$$
\psi_{jk}(x)  = 2^{j/2}\psi(2^{j}x-k)
$$
be the translations and dilations of $\phi$ and $\psi$,
respectively. For any fixed integer $j_0$,
$$
\{\phi_{j_0k},\psi_{jk}|j\geq j_0, k \mbox{ an integer }\}
$$
is an orthonormal basis for $L_2(R)$. Let
$$
\xi_{jk}=\langle f,\phi_{jk} \rangle
$$
and
$$\theta_{jk}=\langle f,\psi_{jk} \rangle
$$
be the usual inner product of a function $f\in L_2(R)$ with the
wavelet basis functions. Then $f$ can be expressed as an infinite
series
\begin{equation}
f(x)=\sum_k\xi_{j_0k}\phi_{j_0k}(x)+\sum_{j=j_0}^{\infty}\sum_k\theta_{jk}\psi_{jk}(x).
\label{eq:series}
\end{equation}
Since the function $f$ is not known, the wavelet coefficients
are estimated using the discrete wavelet transform (DWT). If $f$ is
represented as a vector of dyadic length $n=2^J$ for some positive
integer $J$, then the DWT will provide a total of $n$ estimated
coefficients $\xi_{j_0k}$ and $\theta_{jk}$ over the indices
$j=j_0,j_0+1,\ldots,J-1$ and for all appropriate $k$. The lowest
level possible for $j_0$ is 0, the highest is $J-1$.  If using wavelets periodized to $[0, 1]$, as in this paper, the indices $k$ run from $1$ to $2^j$.

Wavelets have the useful property that they can simultaneously
analyze a function in both time and frequency.  This is done by
projecting the function to be analyzed into several subspaces.  Each
subspace, or resolution level, characterizes a different degree of
smoothness of the function.  The lowest resolution level, associated
with the index $j=j_0$, represents the smoothest or coarsest part of
the function.  Increases in the index $j$ correspond to decreasing
smoothness.  The highest resolution levels $j$ therefore represent
the behavior of the function at the highest frequencies.

Since the wavelet series in Equation~\ref{eq:series} forms an orthogonal
representation, the sum of the projections in these resolution
levels is the function $f$. Because of the compact support of the
wavelet functions $\phi$ and $\psi$, wavelets also provide the
ability to localize the analysis within each subspace. The higher
the resolution, the greater the degree of localization.

By varying the resolution level $j$ wavelets have the ability to
zoom in or out onto the detailed or smooth structure of $f$. This is
referred to as the multiresolution property of wavelets. Changing
the index $k$ allows wavelets to localize the analysis.  These
properties enable wavelets to model functions of very irregular
types, as well as smooth functions.

\section{Estimating Noise within a Profile}
Suppose a sequence of noise contaminated profiles $f^t$ is
observed as in Equation~\ref{eq:model}. The wavelet coefficients of $f^t$
are estimated with the DWT via the noisy observed signal $y^t$
$$
\tilde\theta=n^{-1/2}Wy^t=n^{-1/2}Wf^t+n^{-1/2}W\varepsilon^t
$$
where
$$
\tilde\theta^t=\{\tilde\xi^t_{j_0,1},\tilde\xi^t_{j_0,2},\ldots,\tilde\xi^t_{j_0,2^{j_0}},
\tilde\theta^t_{j_0,1},\tilde\theta^t_{j_0,2},\ldots,\tilde\theta^t_{j_0,2^{j_0}},
\tilde\theta^t_{j_0+1,1},\ldots,\tilde\theta^t_{J-2,2^{J-2}},
\tilde\theta^t_{J-1,1},\ldots,\tilde\theta^t_{J-1,2^{J-1}}\}.
$$
In practice, the factor $n^{-1/2}$ is left out since it is merely a
scale factor that cancels itself out when applying the inverse DWT.
In this paper, the scalar $n^{-1/2}$ will be ignored. Each vector is
partitioned or organized as follows: the first $2^{j_0}$ components
$\tilde\xi_{j_0}$ represents the coarsest, smoothest part of the
function. The next $2^{j_0}$ components $\tilde\theta_{j_0}$ are the
next coarsest part, the next $2^{j_0+1}$ components are the next
coarsest after that, etc. The last $n/2=2^{J-1}$ components
represent the finest details of the original data $y^t$. The
lowest two coarse parts reside at the same resolution level $j_0$.  This is reflected with a notation difference ($\tilde\theta$ vs $\tilde\xi$).

By the orthogonality of the wavelet transform, the estimated wavelet
coefficients $\tilde\theta^t_{j,k}$ are independent normal random
variables whose means are the true, unknown wavelet coefficients
$\theta^t_{j,k}$ and have variance $\sigma^2$.

Two cases are considered in the estimation of the noise in the
observed signal $y^t$.  The first examines functions $f^t$ whose
wavelet coefficients $\theta^t_{J-1}$ at the highest level are all
zero. This is equivalent to saying that $f^t$ exhibits no structure
at the finest detail level in the wavelet domain.
Thus, this essentially assumes  a smoothness constraint on $f^t$.
It should be noted that this condition is dependent on the choice of wavelet basis.

Such an assumption is often not unreasonable.  Common wavelet analysis
typically involves estimating $f^t$ from the observed $y^t$ by
thresholding, or shrinking, the DWT coefficients.  The properties and advantages of differing types of thresholding methods are well documented in the literature.  See \cite{donoho4, donoho, cai, caisilverman, chicken3, chicken4}.
In practice,
thresholding techniques generally result in this
highest level of detail coefficients being set to 0 except in the
cases of functions with high degrees of irregularity (jumps,
for example). Thus, situations in which an estimate of $f^t$ found with thresholding is considered acceptable
would meet the smoothness constraint given above.

Note that merely subtracting the estimate of $f^t$ obtained via
thresholding methods from the observed data $y^t$ will not give a
vector of noise.  This is because thresholding is conservative: it
sets too many of the coefficients to zero, leaving functional
structure in the difference. Thus the difference is not noise, but
contains both variance and functional components of $y^t$.

Assuming that the $f^t$ are smooth as discussed above,
then the
$\tilde\theta_{J-1,k}$ are normal (0, $\sigma^2$).  These $n/2$
random variables can be used to estimate the noise $\varepsilon^t$
without considering contamination by functional structure.  In this case, we can consider the use of the sample variance estimator, even though this estimator is susceptible to outliers.  For these smooth profiles $f^t$, our proposed method will pair well with a simultaneous profile structure monitoring method since the noise monitoring will be unaffected by profile structure changes.

The second case considered is when structure of $f^t$ exists in the
highest detail resolution level of the DWT of the observed profile $y^t$.
Although we are not interested in monitoring the structure in this paper, we must still account for it or our noise monitoring method will be confounded with structure changes.
Any method of variance estimation must account for the structural elements of $f^t$ that are
evident at this level.  To do this, some additional mild assumptions are
required.  The first assumption is that relatively few of the coefficients at
this level have structural elements of $f^t$.  In the simulation section of this paper, we allow up to 30\% of these coefficients to possess structure.
Functional structure at this resolution
level equates to highly irregular features such as discontinuities
or points of non-differentiability.  In such cases, most of the
$\tilde\theta^t_{J-1,k}$ are normal with mean $0$, while a few are
normal with some unknown, nonzero mean.  All have variance
$\sigma^2$.  The second additional assumption is that these nonzero
means are large compared to the noise.  Thus, the functions
considered in this second case include functions that are less
smooth than those in the first case by the inclusion of irregular
features.

To estimate noise in this second case, we use two estimators that are robust to outliers.  These are the median absolute deviation (MAD) and the pseudo-standard error (PSE) of \cite{lenth}. The PSE was devised to detect a few large
factor effects out of many zero effects in the context of factorial
design of experiments problems.  In this paper, Lenth's method is
applied to the wavelet coefficients in the noise subspace.  Most of these coefficients have mean 0 and a few have large means.

In the next section, statistical tests are proposed for detecting
changes in the noise in a sequence of profiles for both cases: functions exhibiting no structure at the highest
detail level and those that do.  We compare all three variance estimates: sample variance, MAD and PSE.  The sample variance is not typically used in wavelet analysis due to its susceptibility to outliers, but we include it in our simulation study for illustrative purposes.

Finally, the sample size $n$ for each profile
has an effect on the ease with which the assumptions above are met.
Larger values of $n$ imply that there are higher resolution levels available in the
DWT of $y^t$.  For example, doubling $n$ gives one additional, high
resolution level of details (i.e., $J$ is increased by $1$).  Two
advantages result because of this. First, since the highest level of
detail coefficients $\theta^t_{J-1}$ is twice as long before, the
estimates of $\sigma_t$ obtained from this vector will be improved.
Second, with more resolution levels available, the amount of
structure appearing in the highest resolution level is decreased.
Therefore, the conditions of smoothness will be more easily met.

\section{Changepoint Methods for Monitoring Noise}
The tests used to determine when the variance in Equation~\ref{eq:model}
changes will be based on maximizing a likelihood ratio.  The
methods we propose will use all profiles observed up to the current index $T$ to determine
not only that a change has occurred, but when the change occurred
and the magnitude of the change.

\subsection{Hypotheses}
The null hypothesis is that for some index $T$,
$$
H_0: \sigma_0=\sigma_1=\cdots =\sigma_T
$$
where $\sigma_0$ is the known in-control value of the noise, and
$\sigma_t$ is the noise from profile $t$. The alternative considered
is that the noise changes immediately after some time $\tau$:
$$
H_a: \sigma_0=\sigma_1=\cdots =\sigma_\tau \neq \sigma_{\tau+1}=\cdots =\sigma_T.
$$
The value for the out-of-control noise is not specified.  The
proposed method will estimate both the change time $\tau$ and the
value of the out-of-control noise.  The form of $f^t$ is immaterial
except for the smoothness constraints discussed in Section 3.

\subsection{Likelihood}

\subsubsection{Median Absolute Deviation}
If there is no structural component in the highest level
of wavelet coefficients $\tilde\theta_{J-1,k}$, these coefficients
can be assumed to follow a Normal(0, $\sigma^2$) distribution.
At this level of detail, there are $n/2$ coefficients where $n$ is the length of the
profile.

For each observe profile $y^t$, we obtain the coefficients
$\tilde\theta_{J-1,k}^t$ via the DWT and estimate the noise
$\sigma_t$ based on the median absolute deviation (MAD) estimator.
$$
s_{M,t}=\mbox{ MAD }(\tilde\theta^t_{J-1})=\mbox{median
}(|\tilde\theta^t_{J-1,k}-\mbox{median}(\tilde\theta^t_{J-1})|)/c
$$
where $c=\Phi^{-1}(3/4)$ and $\Phi$ is the cumulative distribution function of a standard normal random variable.  The constant $c$ is necessary
to create an unbiased estimate of the magnitude of the noise
$\sigma$.
Since the mean and median are 0 by assumption, we simplify this to
$$
s_{M,t}=\mbox{median }(|\tilde\theta^t_{J-1,k}|)/c.
$$

To form the likelihood, the distribution of the $s_{M,t}$ is needed.
The distribution of the $|\tilde\theta^t_{J-1,k}|$ is
$$
F_{1,\sigma}(x)=\Phi(x/\sigma)-\Phi(-x/\sigma), \hspace{.1in} x>0
$$
with density
$$
f_{1,\sigma}(x)=2\sigma^{-1}\phi(x/\sigma), \hspace{.1in} x>0
$$
where $\phi$ is the density for the
standard normal random variable (the use of $\phi$ for the wavelet
functions will be dropped for the remainder of the paper).
The median of the absolute values of these coefficients
is given by mean of the $\frac{n}{4}$ and $\left(\frac{n}{4}+1\right)$ order statistics
$$
s_{M,t}=\mbox{median }(|\tilde\theta^t_{J-1,k}|)/c=
\frac{|\tilde\theta^t_{J-1}|_{(n/4)}+|\tilde\theta^t_{J-1}|_{(n/4+1)}}{2c}
$$

The density of $s_{M,t}$ is then found by integrating a joint density:
\begin{eqnarray*}
f_{M,\sigma}(s)&=&\int_{Y}f_{S,Y}(s,y)dy\\
\\&=&
\int_{0}^{cs}\frac{8c(n/2)!}{\sigma^{2}[(n/4-1)!]^{2}}\phi\left(\frac{y}{\sigma}\right)\phi\left(\frac{2cs-y}{\sigma}\right)
\left[2\Phi\left(\frac{y}{\sigma}\right)-1\right]^{n/4-1}\left[2-2\Phi\left(\frac{2cs-y}{\sigma}\right)\right]^{n/4-1}dy
\end{eqnarray*}
where $S$ = $s_{M,t}$ as above and $Y$ = $|\tilde\theta^t_{J-1}|_{(n/4)}$.
For any $n$, this integral can be approximated numerically for $\sigma$=1 and generalized to any
$\sigma$ via the relation
$$
f_{M,\sigma}(s)=\frac{1}{\sigma}f_{M,1}\left(\frac{s}{\sigma}\right).
$$

The likelihood under the null hypothesis is
$$
L_0=\prod_{t=1}^Tf_{M,\sigma_0}(s_{M,t}).
$$
Under the alternative, the likelihood is
$$
L_a=\prod_{t=1}^{\tau}f_{M,\sigma_0}(s_{M,t})\prod_{t=\tau+1}^{T}f_{M,\sigma}(s_{M,t}),
$$
and the likelihood ratio is then
$$
L_a/L_0=\prod_{t=\tau+1}^T\frac{f_{M,\sigma}\left(s_{M,t}\right)}{f_{M,\sigma_0}\left(s_{M,t}\right)}.
$$
For any value of $\tau$, we can estimate $\sigma$ as
$$
\hat\sigma_M(\tau)=\sigma_{0}\cdot\frac{\frac{1}{T-\tau}\sum_{t=\tau+1}^T\mbox{
MAD }(|\tilde\theta^t_{J-1}|)} {\frac{1}{\tau}\sum_{t=1}^\tau\mbox{
MAD }(|\tilde\theta^t_{J-1}|)}.
$$
The likelihood ratio can then be expressed as a function of $\tau$
$$
h_M(\tau)=\prod_{t=\tau+1}^{T}\frac{f_{M,\hat\sigma_M(\tau)}\left(s_{M,t}\right)}{f_{M,\sigma_0}\left(s_{M,t}\right)}.
$$
This statistic can be implemented as a control chart where, for each newly observed profile $t$,
maximize $h_M(\tau)$ over all $\tau$
and compare this maximum to an upper control limit (UCL).  The value of the
UCL, $UCL_M$, can be determined by simulation to achieve an in-control
average run length (ARL) of 200, for example. A table of estimated UCL values for various run lengths and sample sizes can be found in \cite{m999}.

A change in the noise is signaled when $\max_\tau h_M(\tau)>UCL_M$.
When a change is signaled, the changepoint $\tau$ can be estimated with the value of $\tau$ that maximizes
the likelihood ratio with these estimated values of $\sigma$
substituted.  Thus, the final estimate of $\tau$ is then
$$
\hat\tau=\arg\max_{0\leq\tau<T} h_M(\tau),
$$
and the estimate of $\sigma$ is then $\hat\sigma_M(\hat\tau)$.

\subsubsection{Sample Variance}
For comparison, the noise in each profile is also estimated with the
sample variance $s^2_{V,t}$ rather than the MAD estimator $s_{M,t}$.
In general, this estimate is not the preferred method for estimating noise
with wavelets. The sample variance is less robust to outliers than
the MAD.  For functions with structure (not just noise) in the
highest level of coefficients, the sample variance would be artificially
high.

Under the null, $v_t=(n/2-1)s^2_{V,t}/\sigma^2_0$ is a $\chi^2_{n/2-1}$ random variable.
Then
$$
L_0=\prod_{t=1}^Tf_{\chi^2}(v_t).
$$
Under the alternative,
$$
v_t=\frac{(n/2-1)s^2_{V,t}}{\sigma^2}\cdot\frac{\sigma^2}{\sigma_0^2}\sim
\frac{\sigma^2}{\sigma_0^2}\cdot\chi^2_{n/2-1},
$$
and the likelihood ratio is
$$
L_a/L_0=\prod_{t=\tau+1}^T\frac{\frac{\sigma_0^2}{\sigma^2}f_{\chi^2}\left(\frac{\sigma_0^2}{\sigma^2}v_t\right)}{f_{\chi^2}\left(v_t\right)}.
$$
As before, both $\tau$ and $\sigma$ must be estimated ($\sigma_0$ is
known). The natural counterpart to estimating $\sigma$ in this
instance is
$$
\hat\sigma^2_V(\tau)=\sigma^2_0 \cdot\frac{\frac{1}{T-\tau}\sum_{t=\tau+1}^T
s^2_{V,t}} {\frac{1}{\tau}\sum_{t=1}^{\tau} s^2_{V,t}}
$$
for any $\tau$. The likelihood ratio is then a function of $\tau$
$$
h_V(\tau)=\prod_{t=\tau+1}^T\frac{\frac{\sigma_0^2}{\hat\sigma^2_V}f_{\chi^2}\left(\frac{\sigma_0^2}{\hat\sigma^2_V}v_t\right)}{f_{\chi^2}\left(v_t\right)}.
$$
This, too, can be implemented as a control chart by maximizing over all $\tau$
and comparing this maximum to an UCL.  A change in the noise would be
signaled when $\max_\tau h_V(\tau)>UCL_V$. When a change is signaled,
choose the value of $\tau$ that maximizes the likelihood ratio with
these estimated values of $\sigma$ substituted and the final
estimate of $\tau$ would be
$$
\hat\tau=\arg\max_{0\leq\tau<T} h_V(\tau).
$$
With this approach, the estimate of $\sigma^2$ is then $\hat\sigma^2_V(\hat\tau)$.

\subsubsection{Pseudo-Standard Error}
The third method considered estimates $\sigma$ with Lenth's PSE.  This estimate is defined as
$$
s_{P,t}=1.5\times \mbox{median}|\tilde\theta^*_{J-1,k}|
$$
where the $\tilde\theta^*_{J-1,k}$ are those $\tilde\theta_{J-1,k}$
which have absolute values less than $2.5s_0$ with
$$
s_0=1.5\times \mbox{median}|\tilde\theta_{J-1,k}|.
$$
This estimate assumes there are few non-zero means among the
normal random variables $\tilde\theta_{J-1,k}$, and that these means
are large.

Given $s_0$, the number of random variables $\tilde\theta^*_{J-1,k}$
is not constant from one profile to the next.  Let $N^t$ be the
number of  $\tilde\theta^*_{J-1,k}$ from profile $t$.  Then,
$E(N^t)\approx 0.99\frac{n}{2}$ if there is no structure at this highest level
of coefficients $\tilde\theta_{J-1,k}$, i.e., all coefficients have
mean 0. For odd $N^t$, the density of $s_{P,t}$ given $s_0$, is
\begin{eqnarray*}
f_{P,\sigma}(s)&=&c_t\sigma^{-1}
\frac{\phi(s/(1.5\sigma))}{\Phi(2.5s_0/\sigma)-0.5}
\left[\frac{\Phi(s/(1.5\sigma)-0.5}{\Phi(2.5s_0/\sigma)-0.5}\right]^{\frac{N^t-1}{2}} 
\left[1-\frac{\Phi(s/(1.5\sigma)-0.5}{\Phi(2.5s_0/\sigma)-0.5}\right]^{\frac{N^t-1}{2}}
\end{eqnarray*}
where the constant $c_t$ depends only on $N^t$.

For even $N^t$, the density of $s_{P,t}$ given $s_0$ can only be found numerically by integrating a joint density, as with the MAD estimator. However, due to the fact that this distribution is being conditioned on $s_0$, the integral would have to be computed for each value of $s_0$. Calculating this integral the required number of times would be too computationally expensive. Therefore, we estimate this density by using the closed form density given above. This approximation is close enough to the true density that it does not significantly effect the procedure.

As above, the likelihood ratio is
$$
h_P(\tau)=\prod_{t=\tau+1}^T\frac{f_{P,\hat\sigma_P}\left(s_{P,t}\right)}{f_{P,\sigma_0}\left(s_{P,t}\right)},
$$
where the estimate of $\sigma$ is
$$
\hat\sigma_P(\tau)=\sigma_{0}\cdot\frac{\frac{1}{T-\tau}\sum_{t=\tau+1}^T
s_{P,t}} {\frac{1}{\tau}\sum_{t=1}^\tau s_{P,t}}.
$$
With this approach a change is signaled when $\max_\tau h_P(\tau)>UCL_P$, and the
estimates of the changepoint $\tau$ and $\sigma$ are
$$
\hat\tau=\arg\max_{0\leq\tau<T} h_P(\tau)
$$
and
$$
\hat\sigma_P =\hat\sigma_P(\hat\tau)
$$
respectively.

\section{Implementing the Changepoint Procedure}
The three methods were each tested on three types of functions: functions with no structure at the highest detail level, functions with 1$\%$ structure at the highest detail level, and functions with 5$\%$ structure at the highest detail level. In order to ensure that there would always be at least one structural component present when a function was assumed to contain structural contamination at the top level, the number of coefficients representing structural components was set to
$$
N_s = \left\lceil p\cdot \left(\frac{n}{2}\right)\right\rceil ,
$$
where $p$ is the proportion of coefficients that represent structure. Therefore the actual percentage of structure present in the top level was always greater than the percentage indicated. Again, with the exception of the number of top level wavelet coefficients representing structural contamination, the underlying function is not assumed to follow any specific form.

Two different values were considered for the size of the structural components: $3\sigma\sqrt{2\log (n)}$ and $\sigma\sqrt{2\log (n)}$. The value $3\sigma\sqrt{2\log (n)}$ was chosen to represent functions whose top level structural components are large enough to be clearly differentiated from noise. On the other hand, $\sigma\sqrt{2\log (n)}$ was chosen to represent functions whose top level structural components are approximately at the level at which wavelet thresholding would take place.  For example, the popular VisuShrink estimator \cite{donoho} uses this second, smaller value to classify a wavelet coefficient as noise or structure.  Changing the size of the structural components should have the largest effect on the sample variance based estimator, while having a minimal effect on the PSE based estimator.

Since the form of the underlying functions are of no importance to these methods, random functions with the necessary amount of structure present in the highest detail level were generated and then contaminated with Normal(0, $\sigma^{2}$) noise. Each profile was constructed by first computing the DWT of a zero vector of length $n$. The structural coefficients were then added to the top detail level in locations that were randomly selected for each profile, while the rest of the coefficients in this top level were left as 0. The remaining detail levels were then populated with coefficients coming from a Uniform(-5, 5) distribution. The inverse DWT was then computed before the normal noise was added.

Simulations were conducted for seven different values of $\sigma$, ranging from 0.50 to 2, where the in-control noise was $\sigma_{0} = 1$. Simulations were run for both $\tau$ = 0 (that is, the noise was out of control at the start of profile monitoring, possibly due to misspecification of $\sigma_0$) and $\tau$ = 20 (the noise variance went out of control after 20 observations).  In the instances where $\tau > 0$ and the test produced a false alarm, $\tau$ was updated to the difference between the original $\tau$ and the run length until the false alarm. The simulation was then rerun for the new $\tau$, and this process was repeated as necessary. In all cases, the UCL was set so that the ARL for in control profiles was 200.

Sample sizes considered range from $n = 2^6$ to $n = 2^{10}$, with results remaining consistent over all sample sizes. Additional tables of results for sample sizes that are not presented here can be found in \cite{m999}.

\subsection{$\tau$ = 0}
The simulation results indicate that when no structure is present in the top level, the use of the sample variance to estimate the noise in the profile is superior to using either MAD or PSE in terms of ARL, but all three methods are comparable in terms of estimating $\tau$ and $\sigma$. However, when there is structure present, the fact that the sample variance is the least robust of the three estimators causes its estimate of $\sigma$ to be artificially high, which in turn results in biased estimates of $\tau$. The direction of this bias depends on the number of structural coefficients present, the size of these coefficients relative to the noise coefficients and the value of the out-of-control $\sigma$.

If $\sigma$ is less than  $\sigma_0$, the possibility exists that the sample variance based method will estimate $\sigma$ to be very close to $\sigma_0$, resulting in an upwardly biased estimate for $\tau$. An example of this case can be seen in Table~\ref{n512.2}. When $p = 0.01$ and $\sigma$ = 0.90, the sample variance based method overestimates $\sigma$ just enough that it cannot immediately reject $H_0$, resulting in an average estimate of 3.68 for $\tau$ instead of 0.

If the structure at the top level of wavelet coefficients is large with respect to the noise, and $\tau$ is 0, then the estimate of $\sigma$ may be large enough to cause the control chart to signal that the profile process is out-of-control immediately.
Although this leads to correctly estimating $\tau$ to be 0, it also will result in incorrect estimates for $\sigma$.   An example of this can be seen in the sample variance column of Table~\ref{n512}. When $p$ is set to 0.05, the estimates for $\sigma$ are more than 2.5 times greater than the true $\sigma$ values.

For $\tau$ = 0, the MAD and PSE methods were mostly comparable in terms of ARL, estimation of $\tau$, and estimation of $\sigma$. Notable exceptions can be seen in Tables~\ref{n512.2}~and~\ref{n512} for $p = 0.05$ and $\sigma$ = 0.90. In these cases the MAD based method often produced a biased estimate for $\sigma$ that was close enough to $\sigma_0$ that the control chart statistic could not immediately signal an out of control state.
This resulted in ARLs that are greater than 20 and estimates for $\tau$ greater than 4. Being more robust to structural contamination, the PSE based method estimated $\tau$ much better than the MAD based method in these extreme cases.

\subsection{$\tau >$ 0}
The real benefit of using the PSE instead of the MAD or sample variance becomes apparent when $\tau >$ 0, as in Table~\ref{tau20}.
When there is a large amount of structure present, both the MAD and sample variance estimator approaches become highly influenced while the PSE approach remains relatively unaffected.
Due to the method used here for handling false alarms, the shortcomings of the MAD and sample variance methods are not obvious just by looking at the ARLs and estimates for $\tau$.
Instead, it is helpful to look at the columns that report the proportion of runs with at least one false alarm, and the average number of false alarms per run given there was at least one. Analyzing these columns for the sample variance estimator with structure present indicates that even though the ARL is 1 and the estimate for $\tau$ is 20 for all $\sigma$, this method is still flawed since all of the 100 test runs each had 20 false alarms.
This means that in all instances, the structure in the highest detail level resulted in an upwardly biased sample variance that was large enough to reject $H_0$ immediately, before $\sigma$ actually went out of control. The MAD based method had similar problems when $p$ = 0.05, having at least one false alarm in more than 99$\%$ of the cases. However unlike the sample variance method, the average number of times the MAD based method signaled a false alarm given there was at least one was less than three.
When $p$ = 0.01, the MAD based method experienced an increase in the number of false alarms.
However, this increase was not nearly as dramatic as the increase present in the sample variance method.

While the MAD may compare favorably to the sample variance, the PSE based method proves to be superior to both when $\tau$ = 20. Inspection of Table~\ref{tau20} indicates that the proportion of runs with at least one false alarm remains consistent between 0.05 and 0.09, even with as much as 5$\%$ structure at the highest detail level. Additionally, the average number of false alarms given there was at least one remains consistent throughout.

In order to test just how robust the PSE based method is to structure in the highest detail level, an additional simulation was run with the proportion of structural coefficients set to $p$ = 0.30. The results, which can be seen in Table~\ref{p.30}, are comparable to the results for smaller $p$ in terms of ARL, estimation of $\tau$ and $\sigma$, and the false alarm rate.

These results for the PSE imply it will work well in conjunction with a profile structure monitoring method.  It clearly can monitor the noise independently of most changes within the profile structure.

\subsection{Comparison to Other Methods}

While there are no existing techniques that are directly comparable to the method presented here for monitoring the noise variance (without the need to estimate or assume a known form of $f$), for the sake of completeness we can compare our method to those that simultaneously monitor for changes in the functional portion of the profile along with the noise variance. \cite{ztw08} presented such a method, known as NEWMA, which is based on EWMA control charts. ARL comparisons between our method and the NEWMA chart can be seen in Table~\ref{zou}.

Although the NEWMA chart outperforms our PSE based method in terms of ARL, it is important to acknowledge the fact that the NEWMA chart makes use of the assumption that the underlying in control function is known. Our method does not make this assumption; and aside from the fact that subtracting the known function from an observed profile results in no structural contamination in the top detail level of wavelet coefficients, our method cannot take advantage of this assumption. The benefit of our wavelet based method over others, such as NEWMA, is that it offers a Phase II monitoring scheme for changes in noise variance without the need for any Phase I analysis on the functional portion of the profiles. However, under the assumption that the function is known, the sample variance can safely be used to produce an unbiased estimate of $\sigma$; and as can be seen in Table~\ref{zou}, our method performs comparably to NEWMA in terms of ARL when using the sample variance rather than the PSE.

\subsection{Concluding Remarks}
In this paper we propose a changepoint method for monitoring profile variance with three different variance estimators.  Two of these variance estimators are robust, the MAD and PSE, while the third, the sample variance, is sensitive to outliers.  These three variance estimators are applied to the highest resolution projection level of the profile provided by the DWT.  This application is done within profiles, rather than between them.  The monitoring method uses a likelihood ratio to signal when an out-of-control condition has occurred, provides an estimate of the time the process went out-of-control and estimates the non-nominal variance.

When there is structural contamination present in the top resolution level of wavelet coefficients, the results of our simulation study strongly suggest using the PSE rather than the sample variance or MAD in this application. Being non-robust to outliers, the sample variance will be artificially large with even small amounts of structural contamination present, causing the proposed test to signal an immediate change even if no such change has occurred. This is not unexpected.  In general, the sample variance is not used in wavelet analysis, but was included here for the sake of comparison.

These problems associated with the sample variance at low levels of structural contamination of the wavelet noise projection are also evident with the MAD at higher levels of contamination.  The MAD, which is considered a robust estimator of $\sigma$, is not robust enough for this application.  It outperforms the sample variance, but is inferior to the PSE.

If the profiles being considered are known to come from a class of very smooth functions (no profile structure present in the wavelet noise projection), then any of the three monitoring methods considered here should yield accurate results. However, since only noisy profiles are observed in practice, one should not assume that the underlying profiles are sufficiently smooth enough for sample variance or MAD use.  Additionally, if our proposed method is used in conjunction with a profile structure monitoring method, it is desirable that changes in the function structure occurring in the noise projection are not confounded with changes in the variance.  The PSE method is shown to perform very well in these cases, as well as for smoother profiles, and is therefore the suggested method.

\vspace{2cm}
\noindent
Disclaimer: The views expressed in this article are those of the authors and do not reflect the official policy of the United States Air Force, Department of Defense, or the United States Government.

\clearpage

\newpage

\begin{table}[ht]
\begin{center}
\begin{tabular}{|cc|rrr|rrr|rrr|}
\hline
 \multicolumn{2}{|c|}{} & \multicolumn{3}{c|}{Var} & \multicolumn{3}{c|}{MAD} & \multicolumn{3}{c|}{PSE} \\
\hline
         p &  $\sigma$ &        ARL & $\hat\tau$ & $\hat\sigma$ &        ARL & $\hat\tau$ & $\hat\sigma$ &        ARL & $\hat\tau$ & $\hat\sigma$ \\
\hline
      0.00 &       2.00 &       1.00 &       0.00 &       2.00 &       1.00 &       0.00 &       1.99 &       1.00 &       0.00 &       1.99 \\

           &       1.50 &       1.00 &       0.00 &       1.50 &       1.00 &       0.00 &       1.51 &       1.00 &       0.00 &       1.52 \\

           &       1.25 &       1.03 &       0.00 &       1.25 &       1.59 &       0.01 &       1.28 &       1.67 &       0.10 &       1.29 \\

           &       1.10 &       2.57 &       0.20 &       1.12 &       5.64 &       0.72 &       1.15 &       6.43 &       1.29 &       1.15 \\

           &       0.90 &       2.29 &       0.02 &       0.88 &       5.13 &       0.54 &       0.87 &       6.45 &       0.91 &       0.86 \\

           &       0.75 &       1.00 &       0.00 &       0.76 &       1.23 &       0.01 &       0.74 &       1.58 &       0.02 &       0.73 \\

           &       0.50 &       1.00 &       0.00 &       0.50 &       1.00 &       0.00 &       0.50 &       1.00 &       0.00 &       0.50 \\

           &            &            &            &            &            &            &            &            &            &            \\

      0.01 &       2.00 &       1.00 &       0.00 &       2.13 &       1.00 &       0.00 &       2.03 &       1.00 &       0.00 &       2.01 \\

           &       1.50 &       1.00 &       0.00 &       1.60 &       1.01 &       0.00 &       1.54 &       1.03 &       0.00 &       1.49 \\

           &       1.25 &       1.00 &       0.00 &       1.33 &       1.47 &       0.03 &       1.29 &       1.71 &       0.08 &       1.28 \\

           &       1.10 &       1.27 &       0.00 &       1.19 &       5.03 &       0.61 &       1.15 &       6.35 &       0.90 &       1.15 \\

           &       0.90 &      13.66 &       3.68 &       0.94 &       6.48 &       1.17 &       0.88 &       6.86 &       0.48 &       0.87 \\

           &       0.75 &       1.03 &       0.00 &       0.80 &       1.34 &       0.00 &       0.75 &       1.53 &       0.02 &       0.74 \\

           &       0.50 &       1.00 &       0.00 &       0.53 &       1.00 &       0.00 &       0.51 &       1.00 &       0.00 &       0.50 \\

           &            &            &            &            &            &            &            &            &            &            \\

      0.05 &       2.00 &       1.00 &       0.00 &       2.55 &       1.00 &       0.00 &       2.12 &       1.00 &       0.00 &       2.04 \\

           &       1.50 &       1.00 &       0.00 &       1.91 &       1.00 &       0.00 &       1.59 &       1.01 &       0.00 &       1.52 \\

           &       1.25 &       1.00 &       0.00 &       1.61 &       1.18 &       0.00 &       1.36 &       1.46 &       0.06 &       1.31 \\

           &       1.10 &       1.00 &       0.00 &       1.41 &       2.71 &       0.20 &       1.21 &       4.86 &       0.84 &       1.18 \\

           &       0.90 &       1.55 &       0.00 &       1.16 &      21.29 &       5.92 &       0.91 &       8.04 &       1.27 &       0.88 \\

           &       0.75 &      11.11 &       1.69 &       0.94 &       1.69 &       0.04 &       0.78 &       1.73 &       0.01 &       0.75 \\

           &       0.50 &       1.00 &       0.00 &       0.64 &       1.00 &       0.00 &       0.53 &       1.00 &       0.00 &       0.51 \\
\hline
\end{tabular}
\caption{ARL, $\hat\tau$, and $\hat\sigma$ for the three methods with different p and $\sigma$, n = 512. Size of structural components equals $\sigma\sqrt{2\log (n)}$. 100 runs.} \label{n512.2}
\end{center}
\end{table}

\begin{table}[ht]
\begin{center}
\begin{tabular}{|cc|rrr|rrr|rrr|}
\hline
 \multicolumn{2}{|c|}{} & \multicolumn{3}{c|}{Var} & \multicolumn{3}{c|}{MAD} & \multicolumn{3}{c|}{PSE} \\
\hline
         p &  $\sigma$ &        ARL & $\hat\tau$ & $\hat\sigma$ &        ARL & $\hat\tau$ & $\hat\sigma$ &        ARL & $\hat\tau$ & $\hat\sigma$ \\
\hline
      0.00 &       2.00 &       1.00 &       0.00 &       1.99 &       1.00 &       0.00 &       2.00 &       1.00 &       0.00 &       2.01 \\

           &       1.50 &       1.00 &       0.00 &       1.50 &       1.01 &       0.00 &       1.50 &       1.01 &       0.00 &       1.50 \\

           &       1.25 &       1.04 &       0.01 &       1.24 &       1.49 &       0.02 &       1.28 &       1.62 &       0.12 &       1.29 \\

           &       1.10 &       2.48 &       0.21 &       1.13 &       6.35 &       1.13 &       1.14 &       5.85 &       1.46 &       1.17 \\

           &       0.90 &       2.38 &       0.10 &       0.88 &       5.17 &       0.54 &       0.87 &       7.36 &       1.29 &       0.87 \\

           &       0.75 &       1.00 &       0.00 &       0.74 &       1.33 &       0.02 &       0.74 &       1.55 &       0.01 &       0.73 \\

           &       0.50 &       1.00 &       0.00 &       0.50 &       1.00 &       0.00 &       0.50 &       1.00 &       0.00 &       0.50 \\

           &            &            &            &            &            &            &            &            &            &            \\

      0.01 &       2.00 &       1.00 &       0.00 &       3.05 &       1.00 &       0.00 &       2.04 &       1.00 &       0.00 &       2.01 \\

           &       1.50 &       1.00 &       0.00 &       2.28 &       1.00 &       0.00 &       1.52 &       1.02 &       0.00 &       1.49 \\

           &       1.25 &       1.00 &       0.00 &       1.90 &       1.49 &       0.05 &       1.29 &       1.61 &       0.05 &       1.29 \\

           &       1.10 &       1.00 &       0.00 &       1.67 &       4.60 &       0.69 &       1.16 &       6.73 &       1.18 &       1.14 \\

           &       0.90 &       1.00 &       0.00 &       1.38 &       6.76 &       1.41 &       0.88 &       7.49 &       1.15 &       0.87 \\

           &       0.75 &       1.55 &       0.00 &       1.15 &       1.26 &       0.01 &       0.74 &       1.59 &       0.00 &       0.74 \\

           &       0.50 &       1.00 &       0.00 &       0.76 &       1.00 &       0.00 &       0.50 &       1.00 &       0.00 &       0.50 \\

           &            &            &            &            &            &            &            &            &            &            \\

      0.05 &       2.00 &       1.00 &       0.00 &       5.18 &       1.00 &       0.00 &       2.12 &       1.00 &       0.00 &       2.00 \\

           &       1.50 &       1.00 &       0.00 &       3.88 &       1.00 &       0.00 &       1.60 &       1.02 &       0.00 &       1.53 \\

           &       1.25 &       1.00 &       0.00 &       3.23 &       1.15 &       0.01 &       1.35 &       1.76 &       0.15 &       1.30 \\

           &       1.10 &       1.00 &       0.00 &       2.85 &       2.76 &       0.28 &       1.21 &       5.95 &       0.90 &       1.16 \\

           &       0.90 &       1.00 &       0.00 &       2.33 &      20.26 &       4.59 &       0.91 &       6.35 &       1.00 &       0.86 \\

           &       0.75 &       1.00 &       0.00 &       1.94 &       1.70 &       0.00 &       0.78 &       1.51 &       0.01 &       0.74 \\

           &       0.50 &       1.00 &       0.00 &       1.30 &       1.00 &       0.00 &       0.52 &       1.00 &       0.00 &       0.50 \\
\hline
\end{tabular}
\caption{ARL, $\hat\tau$, and $\hat\sigma$ for the three methods with different p and $\sigma$, n = 512. Size of structural components equals 3$\sigma\sqrt{2\log (n)}$. 100 runs.} \label{n512}
\end{center}
\end{table}

\begin{sidewaystable}[ht]
\begin{center}
\begin{tabular}{|cc|rrrrr|rrrrr|rrrrr|}
\hline
 \multicolumn{2}{|c|}{} & \multicolumn{5}{c|}{Var} & \multicolumn{5}{c|}{MAD} & \multicolumn{5}{c|}{PSE} \\
\hline
         p &  $\sigma$ &  ARL & $\hat\tau$ & $\hat\sigma$  & $\hat{P}$ & N &  ARL & $\hat\tau$ & $\hat\sigma$ & $\hat{P}$ & N & ARL & $\hat\tau$ & $\hat\sigma$ & $\hat{P}$ & N \\
\hline
      0.00 &       2.00 &       1.00 &      20.00 &       2.01 &       0.09 &       1.00 &       1.00 &      20.00 &       2.00 &       0.07 &       1.00 &       1.00 &      20.00 &       2.00 &       0.09 &       1.00 \\

           &       1.50 &       1.00 &      20.00 &       1.50 &       0.05 &       1.00 &       1.00 &      20.00 &       1.50 &       0.02 &       1.00 &       1.00 &      19.99 &       1.51 &       0.07 &       1.00 \\

           &       1.25 &       1.00 &      20.00 &       1.25 &       0.10 &       1.10 &       1.08 &      19.91 &       1.25 &       0.06 &       1.00 &       1.17 &      20.00 &       1.24 &       0.09 &       1.00 \\

           &       1.10 &       1.51 &      19.89 &       1.11 &       0.03 &       1.00 &       3.22 &      20.02 &       1.13 &       0.09 &       1.00 &       3.64 &      20.39 &       1.14 &       0.06 &       1.17 \\

           &       0.90 &       1.51 &      19.88 &       0.89 &       0.06 &       1.00 &       2.95 &      19.95 &       0.88 &       0.10 &       1.10 &       3.92 &      19.67 &       0.89 &       0.07 &       1.14 \\

           &       0.75 &       1.00 &      20.00 &       0.75 &       0.09 &       1.00 &       1.00 &      19.79 &       0.76 &       0.11 &       1.00 &       1.08 &      19.95 &       0.74 &       0.05 &       1.00 \\

           &       0.50 &       1.00 &      20.00 &       0.50 &       0.05 &       1.00 &       1.00 &      20.00 &       0.50 &       0.05 &       1.20 &       1.00 &      20.00 &       0.50 &       0.07 &       1.00 \\

           &            &            &            &            &            &            &            &            &            &            &            &            &            &            &            &            \\

      0.01 &       2.00 &       1.00 &      20.00 &       3.14 &       1.00 &      20.00 &       1.00 &      20.00 &       2.04 &       0.16 &       1.00 &       1.00 &      20.00 &       2.00 &       0.07 &       1.00 \\

           &       1.50 &       1.00 &      20.00 &       2.35 &       1.00 &      20.00 &       1.00 &      19.99 &       1.51 &       0.13 &       1.08 &       1.00 &      19.99 &       1.49 &       0.09 &       1.00 \\

           &       1.25 &       1.00 &      20.00 &       1.97 &       1.00 &      20.00 &       1.09 &      19.83 &       1.27 &       0.12 &       1.00 &       1.12 &      19.84 &       1.26 &       0.09 &       1.11 \\

           &       1.10 &       1.00 &      20.00 &       1.72 &       1.00 &      20.00 &       2.68 &      17.68 &       1.12 &       0.21 &       1.00 &       3.94 &      19.77 &       1.12 &       0.09 &       1.00 \\

           &       0.90 &       1.00 &      20.00 &       1.41 &       1.00 &      20.00 &       3.71 &      20.23 &       0.89 &       0.18 &       1.06 &       3.61 &      19.57 &       0.88 &       0.09 &       1.00 \\

           &       0.75 &       1.00 &      20.00 &       1.18 &       1.00 &      20.00 &       1.00 &      19.99 &       0.76 &       0.16 &       1.19 &       1.04 &      19.95 &       0.75 &       0.07 &       1.00 \\

           &       0.50 &       1.00 &      20.00 &       0.78 &       1.00 &      20.00 &       1.00 &      20.00 &       0.51 &       0.19 &       1.21 &       1.00 &      20.00 &       0.50 &       0.07 &       1.00 \\

           &            &            &            &            &            &            &            &            &            &            &            &            &            &            &            &            \\

      0.05 &       2.00 &       1.00 &      20.00 &       5.41 &       1.00 &      20.00 &       1.00 &      20.00 &       2.15 &       0.99 &       2.51 &       1.00 &      20.00 &       1.99 &       0.05 &       1.20 \\

           &       1.50 &       1.00 &      20.00 &       4.07 &       1.00 &      20.00 &       1.00 &      20.00 &       1.59 &       1.00 &       2.72 &       1.00 &      20.00 &       1.51 &       0.07 &       1.14 \\

           &       1.25 &       1.00 &      20.00 &       3.39 &       1.00 &      20.00 &       1.00 &      19.45 &       1.31 &       0.98 &       2.58 &       1.23 &      19.97 &       1.27 &       0.08 &       1.00 \\

           &       1.10 &       1.00 &      20.00 &       2.98 &       1.00 &      20.00 &       1.42 &      18.45 &       1.16 &       0.99 &       2.73 &       3.40 &      19.39 &       1.13 &       0.08 &       1.00 \\

           &       0.90 &       1.00 &      20.00 &       2.43 &       1.00 &      20.00 &      14.45 &      25.11 &       0.92 &       1.00 &       2.57 &       3.70 &      20.01 &       0.88 &       0.05 &       1.20 \\

           &       0.75 &       1.00 &      20.00 &       2.03 &       1.00 &      20.00 &       1.11 &      20.02 &       0.79 &       0.99 &       2.69 &       1.05 &      20.00 &       0.75 &       0.05 &       1.20 \\

           &       0.50 &       1.00 &      20.00 &       1.35 &       1.00 &      20.00 &       1.00 &      20.00 &       0.53 &       1.00 &       2.48 &       1.00 &      20.00 &       0.50 &       0.05 &       1.00 \\
\hline
\end{tabular}
\caption{ARL, $\hat\tau$, $\hat\sigma$, proportion of runs with at least 1 false alarm, and average number of false alarms given at least 1 for the three methods with different p and $\sigma$, n = 1024, $\tau$ = 20. Size of structural components equals 3$\sigma\sqrt{2\log (n)}$. 100 runs.} \label{tau20}
\end{center}
\end{sidewaystable}

\begin{table}[ht]
\begin{center}
\begin{tabular}{|c|rrrrr|}
\hline
 \multicolumn{1}{|c|}{} & \multicolumn{5}{c|}{PSE}\\
\hline
         $\sigma$ &  ARL & $\hat\tau$ & $\hat\sigma$  & $\hat{P}$ & N\\
\hline
      2.00 &      1.00 &      20.00 &       2.03 &       0.04 &       1.00 \\

      1.50 &      1.00 &      20.00 &       1.52 &       0.10 &       1.00 \\

      1.25 &      1.30 &      19.82 &       1.27 &       0.07 &       1.00 \\

      1.10 &      4.60 &      20.08 &       1.14 &       0.07 &       1.00 \\

      0.90 &      3.96 &      19.81 &       0.89 &       0.06 &       1.00 \\

      0.75 &      1.04 &      19.77 &       0.75 &       0.04 &       1.25 \\

      0.50 &      1.00 &      20.00 &       0.51 &       0.10 &       1.20 \\
\hline
\end{tabular}
\caption{ARL, $\hat\tau$, $\hat\sigma$, proportion of runs with at least 1 false alarm, and average number of false alarms given at least 1 for the PSE method with $p$ = 0.30, $\tau$ = 20, n=1024 and varying $\sigma$. Size of structural components equals 3$\sigma\sqrt{2\log (n)}$. 100 runs.} \label{p.30}
\end{center}
\end{table}

\begin{table}
\begin{center}
\begin{tabular}{|r|ccc|}
\hline
           & {\bf NEWMA} &  {\bf PSE} &  {\bf VAR} \\
\hline
{\bf $\sigma$=1.1} &       4.14 &      10.83 &       4.73 \\

{\bf $\sigma$=0.7} &       1.43 &        2.1 &       1.01 \\
\hline
\end{tabular}
\caption{ARL comparisons of NEWMA and the PSE and sample variance based wavelet methods for $\sigma$ = 1.1 and 0.7; n = 256, $\sigma_0$ = 1, $\tau$ = 0. For NEWMA c = 2, $\lambda$ = 0.2. 1000 runs.} \label{zou}
\end{center}
\end{table}

\clearpage

\newpage

\noindent {\it Disclaimer}: The views expressed in this article are those of the authors and do not reflect the official policy or position of the United States Air Force, Department of Defense, or the U.S. Government.

\bibliographystyle{natbib}
\bibliography{all}

\end{document}